\documentclass[%
 reprint,
 amsmath,amssymb,
 prb,
 floatfix,
]{revtex4-2}

\usepackage{graphicx}
\usepackage{dcolumn}
\usepackage{bm}
\usepackage{mathtools}
\usepackage{bm}
\usepackage{graphicx}
\usepackage{dcolumn}
\usepackage{bm}
\usepackage{epstopdf}
\usepackage{xcolor}
\usepackage{xspace}
\usepackage[T1]{fontenc}
\usepackage{lmodern} 

\usepackage{hyperref}
\usepackage[all]{hypcap}
\hypersetup{colorlinks, linkcolor=blue,          
    citecolor=blue,        
    filecolor=blue,      
    urlcolor=blue           
}

\begin{document}

\preprint{APS/123-QED}

\title{Magnetic phase diagram and spin Hamiltonian of antiferromagnet Cs$_2$CoI$_4$}

\author{S.D. Nabi$^1$}
 \email{nabid@ethz.ch}
\author{L. Facheris$^1$}%
\author{V. Romerio$^1$}
\author{V. Kocsis$^2$}%
\author{K. Yu. Povarov$^3$}%
\author{D. Sheptyakov $^4$}%
\author{J. Lass$^4$}%
\author{D.G. Mazzone$^4$}%
\author{H. Kikuchi$^5$}%
 \altaffiliation{Present address: Neutron Scattering Division, Oak Ridge National Laboratory, Oak Ridge, Tennessee 37831, USA}
\author{T. Masuda$^5$}%
\author{S.A. Barnett$^6$}%
\author{D.R. Allan$^6$}%
\author{Z. Yan$^1$}%
\author{S. Gvasaliya$^1$}%
\author{A. Zheludev$^1$}%
 \email{http://www.neutron.ethz.ch/}
\affiliation{%
 $^1$Laboratory for Solid State Physics, ETH Zurich, 8093 Z\"urich, Switzerland\\
  $^2$ Leibniz Institute for Solid State and Materials Research, IFW Dresden, 01069 Dresden, Germany\\
 $^3$Dresden High Magnetic Field Laboratory (HLD-EMFL) and W\"urzburg-Dresden Cluster of Excellence ctd.qmat, Helmholtz-Zentrum Dresden-Rossendorf (HZDR), 01328 Dresden, Germany\\
 $^4$PSI Center for Neutron and Muon Sciences, Paul Scherrer Institute, 5232 Villigen, Switzerland\\
 $^5$ Institute for Solid State Physics, University of Tokyo, Kashiwa, Chiba 277-8581, Japan\\
 $^6$  Diamond Light Source, Harwell Science and Innovation Campus, Didcot, Oxfordshire OX11 0DE, United Kingdom}%
 
\date{\today}

\begin{abstract}
We report comprehensive thermodynamic and neutron scattering measurements on the $S$ = 3/2 antiferromagnet Cs$_2$CoI$_4$, a member of the thoroughly studied family of frustrated magnets Cs$_2MX_4$ ($M$ = Cu, Co, Ru, $X$ = Br, Cl, I, O). Unlike previously studied members, Cs$_2$CoI$_4$ undergoes a structural phase transition, for which we determine the low-temperature crystallographic structure. The resulting symmetry reduction strongly affects both the magnetic exchange interactions and single-ion anisotropy. Despite the large parameter space, we propose a minimal magnetic Hamiltonian that reasonably captures the observed excitation spectrum, analyzed using extended SU(4) linear spin-wave theory.

\end{abstract}

\maketitle


\section{\label{sec:level1}Introduction\protect\\}

The title compound belongs to the well-known family of frustrated magnets Cs$_2MX_4$ ($M$ = transition metal, $X$ = halogen or oxygen), which has long served as a platform for exploring low-dimensional and frustrated magnetism. Depending on the choice of magnetic ion and ligand, these materials feature a wide range of possible frustrated exchange pathways manifesting a variety of complex magnetic properties.

A prominent example is the thoroughly studied $S$ = 1/2 compound Cs$_2$CuCl$_4$, which realizes a distorted triangular lattice model connecting the physics of one-dimensional (1D) spin chains with the two-dimensional (2D) triangular lattice \citep{CCC1, CCC2, CCC3, CCC4}. Its close analog Cs$_2$CuBr$_4$ exhibits a cascade of 
field-induced phase transitions and magnetization plateaus \citep{CCB1, CCB2}. In contrast, the Co-based members with $S$ = 3/2 that feature single-ion anisotropy were found to be more appropriately described by (quasi)-1D models. These include XXZ-chain material Cs$_2$CoCl$_4$ \citep{CCoC1, CCoC2, CCoC3} and frustrated zigzag-ladder system Cs$_2$CoBr$_4$ \cite{Povarov2020, Facheris2022, Facheris2023, Facheris2024}. The latter also hosts one of the most spectacular observations of a Zeeman ladder hierarchy of spinon bound states \cite{Facheris2023}. More recently, the three-dimensional (3D) $S$ = 1 compound Cs$_2$RuO$_4$ was shown to feature a frustration of single-ion anisotropy planes resulting in a spin-flop-like transition accompanied by a quantum critical point (QCP) \citep{CRO}.

An additional member of this family, Cs$_2$CoI$_4$ ($S$ = 3/2), has so far not been investigated from a magnetic perspective. Unlike the previously mentioned compounds, it goes through a structural phase transition at $T_s \sim$ 51~K. The resulting low-temperature crystal structure had not been solved to date \citep{CCIPowder, CCIDielectric}.

In this work, we report a detailed experimental study of the low-temperature crystal structure, as well as single crystal studies of the magnetic phase diagram and excitation spectrum of Cs$_2$CoI$_4$. Despite the large parameter space permitted by the reduced symmetry, we show that the main features of the excitation spectrum can be captured by a minimal model consisting of two inequivalent zigzag ladders.

\section{\label{sec:level1}Methods\protect\\}

Dark green crystals of Cs$_2$CoI$_4$ were grown by the Bridgman-Stockbarger method. The precursors CsI and CoI$_2$ were stoichiometrically mixed in a glassy carbon crucible sealed inside a quartz tube. 

Single crystal samples of Cs$_2$CoI$_4$ were characterized in a series of magnetic and thermodynamic experiments. Magnetic susceptibility was measured using a Quantum Design (QD) Magnetic Property Measurement System (MPMS) SQUID Magnetometer, with a probing field of $\mu_0 H$ = 0.1~T along the principal crystallographic directions, as indexed in the high temperature structure, in a temperature range from 1.8 to 300~K. Magnetization data along the \textbf{a} and \textbf{b} axes were measured with several probes. Between 2 - 12~K and 0 - 14~T these were collected on a 2.5~mg sample using a QD vibrating sample magnetometer (VSM) Physical Property Measurement System (PPMS) insert. Low-temperature magnetization (at 200~mK and 2~K up to 14~T) was measured on a 0.8~mg sample using a Faraday-balance capacitive magnetometer \citep{fb}. Using the same set-up, magnetic torque was measured in fields up to 14~T with temperatures ranging from 200~mK to 4~K. The torque signal corresponds to the deflection of a miniature cantilever on which the sample is mounted. The magnetic field sweep rate was optimized to minimize eddy current heating. Additionally, high field magnetization curves were collected using a compensated pickup coil setup in pulsed fields up to 35~T at the High Magnetic Field Laboratory in Dresden (HLD-EMFL) \cite{HMFL} in a 3-He cryostat reaching a base temperature of 500~mK on a 2~mm$^3$ crystal. Both low-temperature and high-field magnetization data were calibrated to absolute units using those collected on the VSM. Heat capacity was measured on a 0.4~mg sample using a QD PPMS in conjunction with a $^3$He--$^4$He dilution refrigerator (DR) insert. Data were collected using the standard relaxation method. The experimental temperature range spanned 100~mK to 10~K, and magnetic fields up to 14~T. The field was applied along two orthogonal crystallographic directions \textbf{a} and \textbf{b}. Magnetostriction measurements with $\Delta L$ $||$ \textbf{c} and \textbf{b} were performed on single crystals with thickness $L$ = 1.04 mm and 1.42 mm, respectively. The former were performed at the Leibniz Institute for Solid State Materials Research in Dresden with \textbf{H} $||$ \textbf{b}. Together with a PPMS, it allowed for measurements from 1.8~K upwards and up to 9~T. The sample dilation was determined with a miniature capacitive dilatometer \cite{dila} in combination with an Andeen-Hagerling 2700A bridge operated at 1~kHz. Dilation measurements along [010] were performed at ETH Z\"urich with \textbf{H} $||$ \textbf{a}, \textbf{b} with the same capacitance bridge and operated at 1.11~kHz. Here, a PPMS and DR insert were used, allowing for measurements upwards from 200~mK and in magnetic fields up to 14~T. 

High-resolution single-crystal x-ray diffraction (XRD) experiments were carried out on the I19 beamline (EH 1) \cite{I19} at the Diamond Light Source. The Helix cryostat was employed, offering a temperature range down to 30~K. Approximately spherical crystals of roughly 50~$\mu$m in diameter were selected to achieve isotropic absorption effects. Measurements were performed with an incident photon energy of 17.9976~keV, using the Zr K edge, at temperatures of 300 and 30~K. For both temperatures, a detector distance of 160~mm was used. For 30~K, an additional detector distance of 300~mm was employed. This enabled capturing Bragg peaks at higher 2$\theta$, to improve the spatial resolution of the diffraction. A sequence of rocking and azimuthal scans was performed with a counting time of 0.2~s/degree at a transmission of approximately 0.03 – 0.05~\%. The collected peaks were integrated using CrysAlisPro \cite{Crysalis} and the structure determined with SHELX \cite{Shelx}.

\begin{table}[t]
\caption{\label{tab:tablecrys}%
Crystal structural parameters for Cs$_2$CoI$_4$ determined at room temperature using single crystal x-ray diffraction.
}
\begin{ruledtabular}
\begin{tabular}{l|c|c|c|c}
\textrm{Atom} & \textrm{$x$} & \textrm{$y$} & \textrm{$z$} & \textrm{$U_{\text{eq}}$} \\
\colrule
Cs~1 & 0.47695(6) & 0.25     & 0.33221(5) & 0.0506(2) \\
Cs~2 & 0.13325(8) & 0.25     & 0.60469(9) & 0.0803(4) \\
Co & 0.73523(11) & 0.25     & 0.57727(9) & 0.0351(3) \\
I~1  & 0.81339(7)  & 0.25     & 0.40783(6)  & 0.0643(3)  \\
I~2  & 0.49769(7)  & 0.25     & 0.59930(6)  & 	0.0619(3)  \\
I~3  & 0.82652(5)  & 0.50373(6) & 0.65502(6)  & 0.0699(3)  \\
\end{tabular}
\end{ruledtabular}
\end{table}

Neutron powder diffraction experiments were conducted on the HRPT beamline \cite{HRPT} at the Paul Scherrer Institute (PSI). 6~g of finely crushed single crystals of Cs$_2$CoI$_4$ was packed into a vanadium canister and mounted in a regular orange cryostat. Measurements were performed at 70 and 40~K, above and below the structural transition, respectively, with an incoming neutron wavelength of $\lambda$ = 2.45~\AA ~in high resolution mode. The pattern was measured in the range from 0 to 160$^{\circ}$ in 2$\theta$ with a step-size of 0.05$^{\circ}$. Rietveld refinement \cite{Rietveld} was conducted using the FullProf \cite{Fullprof} software package.

Neutron spectroscopy experiments were performed on a 600~mg single crystal of Cs$_2$CoI$_4$ at the multiplexing spectrometer CAMEA \citep{CAMEARSI} at PSI. The sample was installed with the $(0, k, l)$ scattering plane horizontal and mounted in an 11~T vertical magnet with a DR insert. Datasets were collected at a base temperature of $T$ = 100~mK. Seven measurement series were done using fixed incoming neutron energies of $E_i$ = 5.1, 5.23, 6.1, 6.19, 6.23, 6.26, 6.7, and 6.83~meV (elastic resolution $\sim$~0.19~meV) in zero field. At 3 and 6~T, the same energies were measured except 6.19 and 6.26~meV. Sequential steps of approximately 0.08/0.13~meV were used to interlace the datasets and suppress horizontal intensity artifacts. Each incoming energy was measured at 2$\theta$ = -41$^{\circ}$, -45$^{\circ}$. For each 2$\theta$, the sample was rotated over a 110$^{\circ}$ range in 1$^{\circ}$ steps, counting 60 seconds per angle. Data reduction was performed using the MJOLNIR \citep{MJOLNIR} software package. 

Additional neutron spectroscopy measurements were done on a 800~mg single crystal at the multiplexing spectrometer HODACA at JRR-3. 24 integrated analyzers cover a 2$\theta$ range of 46$^{\circ}$. The instrument exhibits a location-dependent energy resolution ranging from 0.13 - 0.21~meV \cite{HODACA}. The sample was installed in the ($h$,$k$,0) scattering plane and mounted in  a $^3$He cryostat. Data were collected at a base temperature of $T$ = 700~mK. The energy transfer was measured in the range $\hbar \omega$ = 1.2 - 3.24~meV in steps of 0.06~meV. For each energy, the sample was rotated over 110$^{\circ}$ in steps of 2$^{\circ}$, counting for 3 minutes per angle. The spectra were analyzed using the \texttt{SUNNY.JL} software package \cite{sunny}.

\begin{figure}[t]
\includegraphics[clip, trim=0.0cm 0.0cm 0.0cm 0.0cm ,width=\linewidth]{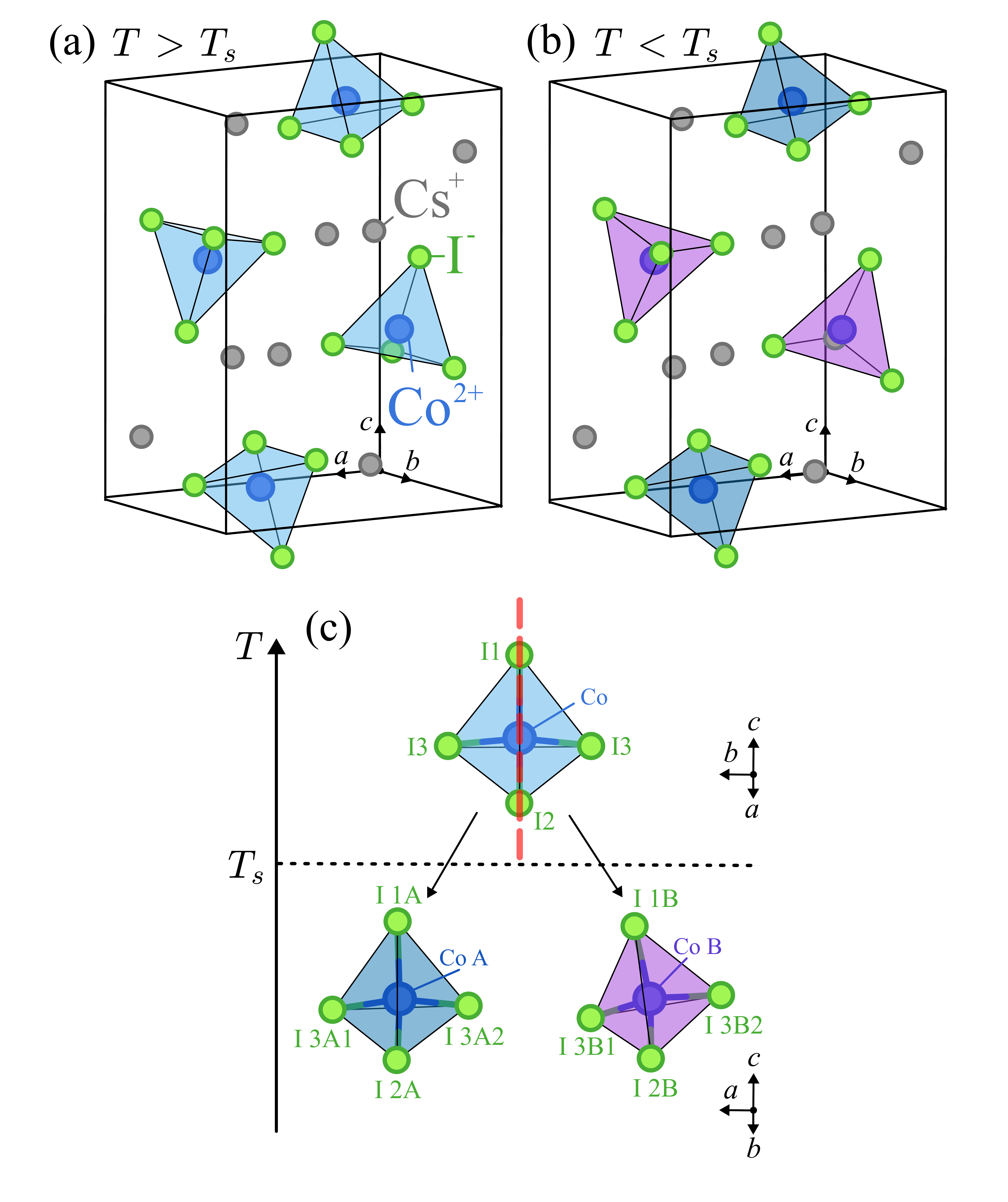}
\caption{\label{fig:crysstruc} Schematic overview of the (a) room temperature unit cell and (b) low temperature unit cell. I$^{-}$ belonging to tetrahedra outside the shown unit cell have been omitted for visibility. (c) shows a zoom-in on the CoI$_4$ tetrahedra at room temperature (top) and two pairs of inequivalent ones below $T_s$ (Co~A left and Co~B right).}
\end{figure}

\section{\label{sec:level1}Experimental results\protect\\}

\subsection{\label{sec:level2}Crystal structure and phase transition}

The room temperature structure of Cs$_2$CoI$_4$ was previously reported in Ref. \cite{CCIPowder}. It crystallizes in an orthorhombic structure, space group $Pnma$ (No. 62), and the lattice parameters are $a = 10.8057(2)$, $b = 8.2681(2)$, and $c = 14.3873(3)$~\AA. This structure was validated using single-crystal XRD on a Bruker APEX-II diffractometer. The refinement with anisotropic thermal factors (quality factor $R = 3.6$~\%) was based on the analysis of 1716 independent Bragg reflections. The results are summarized in Table~\ref{tab:tablecrys} and consistent with the previous study. A schematic overview of the crystal structure at room temperature is given in Fig.~\ref{fig:crysstruc}(a). 

\begin{figure}[b]
\includegraphics[clip, trim=0.0cm 0.0cm 0.0cm 0.0cm ,width=\linewidth]{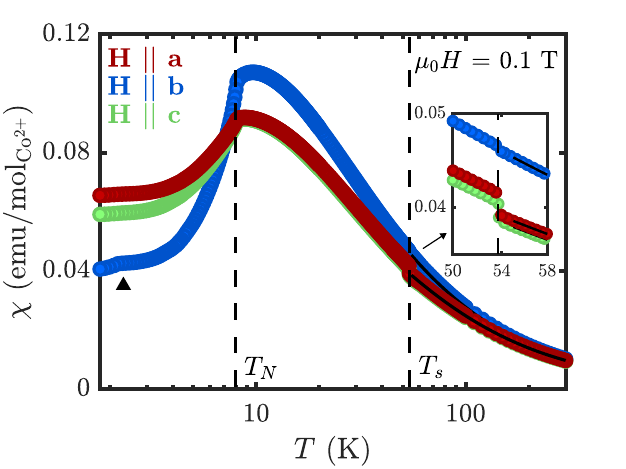}
\caption{\label{fig:susceptibility} Magnetic susceptibility of Cs$_2$CoI$_4$ single crystal in applied fields of $\mu_0H$ = 0.1 T along the three crystallographic directions (symbols). The response along \textbf{a} and \textbf{c} is almost indistinguishable at high temperatures. The solid lines indicate fits to the data as described in the text. Dashed lines (from left to right) show positions of anomalies related to the Néel temperature $T_N$ and structural transition temperature $T_s$. Right inset: the same curves zoomed in between 50 and 58~K from 0.035 to 0.05~emu/mol$_\mathrm{Co^{2+}}$.}
\end{figure}

Previous dielectric and low temperature powder x-ray diffraction experiments indicate a first-order structural phase transition at $T_s$ $\sim$ 51~K \cite{CCIDielectric, CCIPowder}. We confirmed this through anomalies observed in susceptibility [Fig.~\ref{fig:susceptibility}] and sample dilation measurements [Fig.~\ref{fig:dila}].

\begin{figure}[t]
\includegraphics[clip, trim=0.0cm 0.0cm 0.0cm 0.0cm ,width=\linewidth]{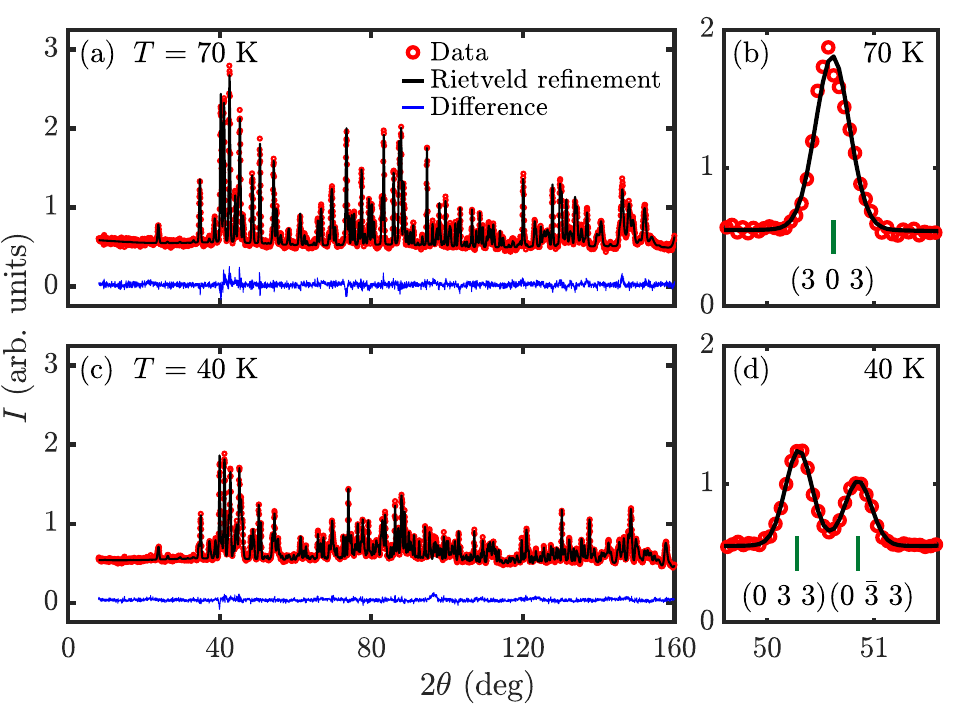}
\caption{\label{fig:HRPT} Neutron powder diffraction pattern at (a) $T$ = 70~K and (c) $T$ = 40~K.  Red circles show the measured data, black lines indicate the fits from Rietveld refinements, and blue lines indicate the difference between the measured data and fits. (b) and (d) show a restricted 2$\theta$ range from (a) and (c), respectively. The vertical green lines represent the peak positions indicated by the accompanying ($h$ $k$ $l$) index.}
\end{figure}

As shown in the following, the low temperature crystal structure was challenging to determine. For this reason, both high-resolution single-crystal XRD and neutron powder diffraction were used in a complementary approach. The XRD measurement at room temperature was consistent with the results in Table~\ref{tab:tablecrys} ($R_1$ = 3.6~\%), confirming the structural model and the sample quality. At 30~K, below $T_s$, additional peaks appeared at forbidden positions, as expected for a lower symmetry group, along with peak splitting indicative of structural domains. Integration suggested a triclinic unit cell below $T_s$. However, the quality factor remained high ($R_\text{int} >$ 30~\%), likely due to difficulties in assigning the present crystallographic domains. Nonetheless, a solution was obtained in the $P_{-1}$ (No. 2) space group. This solution is remarkably similar to the room temperature structure, with only a small distortion of the unit cell and minor adjustments to atomic positions. The high $R_\text{int}$ value propagated into a correspondingly high $R_1 >$ 30~\%. This preliminary solution was subsequently used to further refine the structure against neutron powder diffraction data, where there is no challenge of assigning domains.

Measured powder neutron diffraction patterns at $T$ = 70~K and 40~K are illustrated in Figs.~\ref{fig:HRPT}(a) and (c), respectively. The high-temperature structure was confirmed at $T$ = 70~K. Clear splitting of Bragg peaks is observed below $T_s$. For example, peak (3 0 3) at $T$ = 70~K splits into two peaks at $T$ = 40~K as shown in Figs.~\ref{fig:HRPT}(b) and (d). These splittings could not be accounted for by a phase transition to any symmetry higher than triclinic.

The preliminary XRD low temperature triclinic crystal structure was refined using the Rietveld method with an overall thermal factor (including V impurity) ($R$ = 6.4~$\%$, $\chi^2$ = 2.04). The lattice parameters are $a$ = 8.1432(1), $b$ = 10.7490(2), $c$ = 14.3270(2)~\AA, $\alpha$ = 89.3844(4)$^{\circ}$, $\beta$ = 89.4970(5)$^{\circ}$, $\gamma$ = 89.5376(5)$^{\circ}$. The atomic positions are summarized in Table~\ref{tab:tablecrys2}. The structural transition splits previously equivalent ions into inequivalent A/B pairs (only I~3 splits into four). The low temperature structure is schematically drawn in Fig.~\ref{fig:crysstruc}(b). Note that the structures below and above $T_s$ are very similar, but that the \textbf{a} and \textbf{b} axes are switched by triclinic convention. 

In the following, a multitude of experimental data will be shown. In many of these, measurements were performed both above and below the phase transition. For this reason, the crystallographic axes will be conveniently shown in the orthorhombic notation.

\begin{table}[t]
\caption{\label{tab:tablecrys2}%
Crystal structural parameters for Cs$_2$CoI$_4$ determined at $T$ = 40~K with neutron powder diffraction with an overall thermal factor $U_{\mathrm{ov}}$ = 0.013. Note that the \textbf{a} and \textbf{b} axes were switched compared to the orthorhombic structure due to triclinic convention of the lattice parameters, this is directly reflected in the $x$ and $y$ coordinates compared to Table \ref{tab:tablecrys}.}
\begin{ruledtabular}
\begin{tabular}{l|c|c|c}
\textrm{Atom} & \textrm{$x$} & \textrm{$y$} & \textrm{$z$}  \\
\colrule
Cs~1A & 0.2256(8)     & 0.4735(6) &  0.3353(4) \\
Cs~1B & 0.7492(8)     & 0.0169(6) &  0.8291(4)\\
Cs~2A & 0.2697(8)     & 0.1323(6) &  0.6281(5)\\
Cs~2B & 0.2440(8)     & 0.6286(6) &  0.9074(4)  \\
Co~A & 0.263(2) &  0.236(1)    & 0.924(1)  \\
Co~B & 0.268(2) & 0.739(1)     & 0.579(1)  \\
I~1A  & 0.2711(8)  & 0.3196(7)     & 0.0920(4)   \\
I~1B  &  0.2047(8)  & 0.8116(6) & 0.4103(4)    \\
I~2A  & 0.2717(8)  & 0.9936(6) & 0.9054(5) \\
I~2B  & 0.2736(8)  & 0.4937(6)     & 0.6016(4)  \\
I~3A1  & 0.0030(8)  & 0.3221(6)     & 0.8513(5)    \\
I~3A2  & 0.5111(9)  & 0.3247(7) & 0.8383(5)    \\
I~3B1  & 0.5486(8)  & 0.8349(6)     & 0.6262(4)    \\
I~3B2  & 0.0393(8)  & 0.8185(6) & 0.6863(5)   \\

\end{tabular}
\end{ruledtabular}
\end{table}

The magnetism in Cs$_2$CoI$_4$ stems from the Co$^{2+}$ ions ($S$ = 3/2). At room temperature, there are four equivalent cobalt sites (Co in Table~\ref{tab:tablecrys}). The local environment is constructed by a distorted tetrahedron of iodide anions containing a mirror plane symmetry in the crystallographic $ac$ plane [Fig.~\ref{fig:crysstruc}(c) top]. At low temperatures, there are two inequivalent Co$^{2+}$ ions (Co~A and Co~B) where the local electrostatic environment is additionally distorted, and the mirror plane symmetry is lost. These are illustrated at the bottom of Fig.~\ref{fig:crysstruc}(c). The local environment of Co~A is least impacted and remains almost identical to that of room temperature. Co~B is more significantly affected. Since there is no point group symmetry, the orbital angular momentum is quenched, making the $S$ = 3/2 degrees of freedom appropriate to describe the magnetism of this material. The reduction of overall and point group symmetry potentially has a profound effect on magnetic properties such as the single-ion anisotropy and allowed magnetic exchange interactions. These will be discussed in detail in Sec.~\ref{sec:ModelHamiltonian}.

To prevent multi-domain effects, all thermodynamic properties discussed below (except high-field magnetization and magnetic susceptibility) have been measured after slow cooling over the structural phase transition at 14~T from 60 to 40~K with a cooling rate of 0.1~K/min. This is explained in more detail in the Appendix.

\subsection{\label{sec:level2} Susceptibility and Magnetization}

\begin{figure}[t]
\includegraphics[clip, trim=0.0cm 0.0cm 0.0cm 0.0cm ,width=\linewidth]{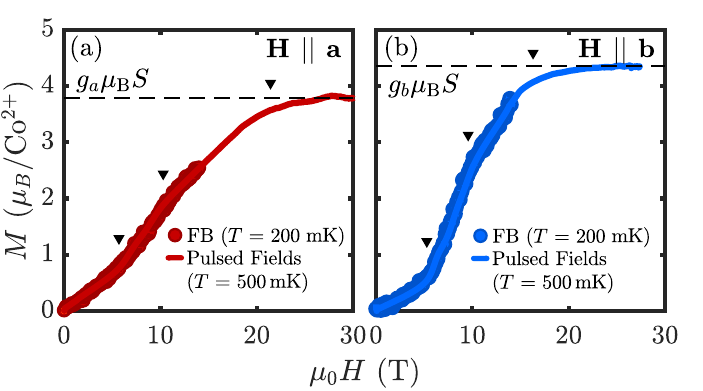}
\caption{\label{fig:LT_mag} Magnetization as a function of applied magnetic field along the (a) \textbf{a} direction and (b) \textbf{b} direction. Circles are data points measured with the Faraday balance (FB) at $T$ = 200~mK, and lines are pulsed field magnetization measurements at $T$ = 500~mK. Dashed black lines indicate the saturation magnetization values. The black triangles indicate anomalies in the magnetization curve.}
\end{figure}

\begin{figure}[t]
\includegraphics[clip, trim=0.0cm 0.0cm 0.0cm 0.0cm ,width=\linewidth]{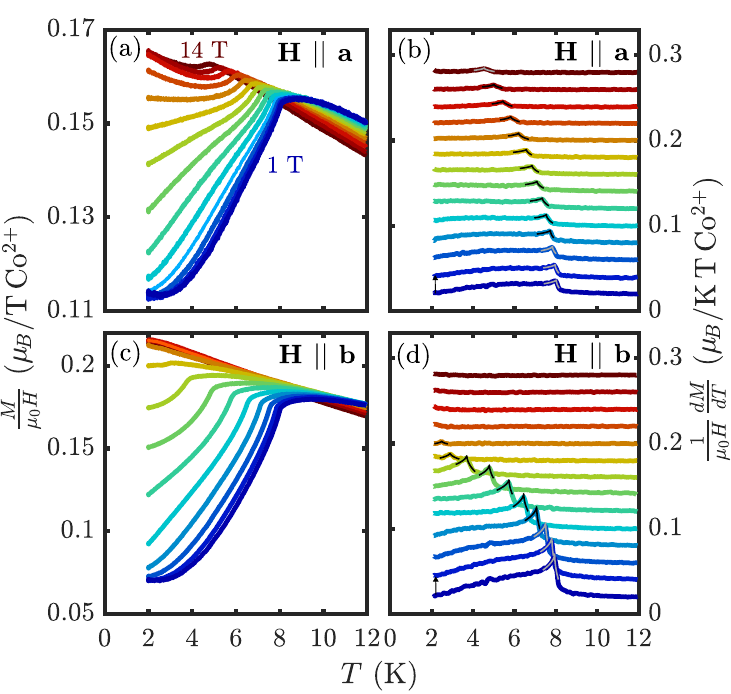}
\caption{\label{fig:VSM} Magnetization as function of temperature at several applied magnetic fields, from 1 to 14~T in steps of 1~T, along the (a) \textbf{a} direction and (c) \textbf{b} direction. Note that the y-axes do not start at zero, for visual purposes. Sub-plots (b) and (d) correspond to the derivative of the magnetization along the \textbf{a} and \textbf{b} directions, respectively, with respect to temperature. In (b) and (d) the curves at each consecutive field are vertically offset by 0.02~$\mu_\mathrm{B}$~K$^{-1}$T$^{-1}$ per Co$^{2+}$ ion. The black and gray lines in (b) and (d) correspond to a phenomenological fit to extract the peak positions of the anomalies.}
\end{figure}

Figure~\ref{fig:susceptibility} shows magnetic susceptibility measured upon warming in a probing field of $\mu_0H = 0.1$~T along all three crystallographic axes. Antiferromagnetic order is observed at $T_N = 8$~K, with an additional anomaly near the structural transition at $T_s = 54$~K (inset). The influence of the structural transition is more pronounced for fields along the \textbf{a} and \textbf{c} directions than along \textbf{b}.

A Curie-Weiss analysis above $T_s$ (55 – 300~K), including a constant background $\chi_0$, yields $g$-factors: $g_{a,c}$ = 2.5(1), $g_b$ = 2.6(1), and Weiss temperatures $\theta_{a,c} = -25.0(1)$~K and $\theta_b = -15.6(1)$~K, indicating antiferromagnetic interactions with moderate frustration $\theta$ $>$ $T_N$ and a degree of anisotropy. The magnetic properties likely change substantially below the structural transition, so the low-temperature properties may not be directly inferred from the high-temperature fits, and these only serve as a reference.

Below $T_N$, a pronounced anisotropy emerges. The susceptibility along \textbf{b} drops fastest, identifying it as the easiest axis. The \textbf{c} axis is slightly easier than \textbf{a}. A minor feature near 2~K is also visible. The lack of local symmetry for the two inequivalent Co$^{2+}$ sites implies that the true easy and hard axes do not necessarily need to align with principal crystallographic directions and can differ on each unique site, in contrast to related compounds such as Cs$_2$CoBr$_4$ and Cs$_2$RuO$_4$ \cite{CRO,Povarov2020}.

Magnetization curves against field along the \textbf{a} and \textbf{b} directions at $T$ = 200~mK (Faraday balance) and  $T$ = 500~mK (pulsed field) are shown in Fig.~\ref{fig:LT_mag}(a) and (b), respectively. For field along \textbf{a}, weak slope changes appear at 8 and 11~T, with saturation at $\mu_0H_{sat,a}$ $\sim$ 22~T ($g_a S = 3.8 \, \mu_{\mathrm{B}}$). Along \textbf{b}, anomalies at 6 and 10~T are more pronounced, saturating at $\mu_0 H_{\mathrm{sat},b} \sim 16 $~T ($g_b S = 4.35 \, \mu_{\mathrm B} $). The extracted $g$-factors are $g_a$ = 2.5 and $g_b$ = 2.9, confirming the magnetic anisotropy.

Magnetization as a function of temperature under fields 1 – 14~T along the \textbf{a} and \textbf{b} directions is shown in Figs.~\ref{fig:VSM}(a) and (c), respectively. These curves show a stronger suppression of $T_N$ with increasing field along \textbf{b} compared to \textbf{a}. The transition temperatures were extracted from phenomenological fits on anomalies in the numerical derivatives [Figs.~\ref{fig:VSM}(b,d)].

\begin{figure}[t]
\includegraphics[clip, trim=0.0cm 0.0cm 0.0cm 0.0cm ,width=\linewidth]{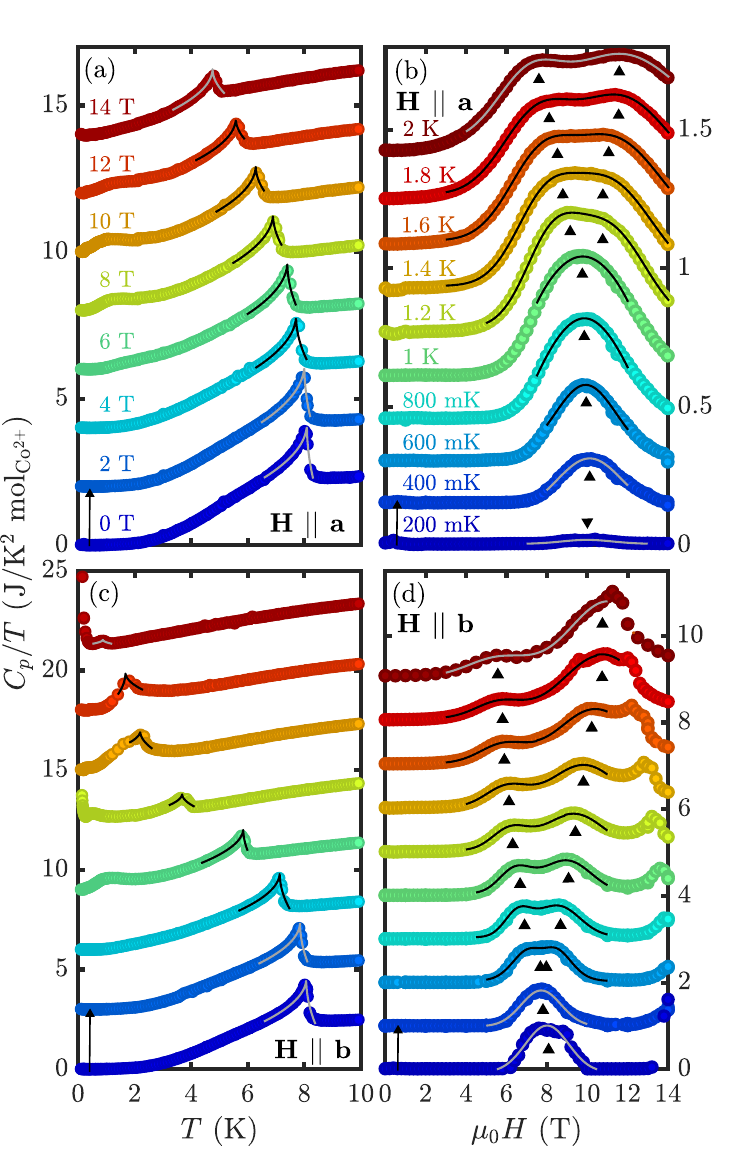}
\caption{\label{fig:HC}Specific heat measured in Cs$_2$CoI$_4$ in a magnetic field applied along two crystallographic directions. (a) and (c) are temperature scans at sequential fields and (b) and (d) field scans at sequential temperatures. Symbols indicate the positions of anomalies in the field scans. Sequential temperature scans are offset by 3~JK$^{-2}$mol$_{\mathrm{Co^{2+}}}^{-1}$ and field scans with 0.15~JK$^{-2}$mol$_{\mathrm{Co^{2+}}}^{-1}$ for \textbf{H} $||$ \textbf{a} and 1~JK$^{-2}$mol$_{\mathrm{Co^{2+}}}^{-1}$ for \textbf{H} $||$ \textbf{b}. Black and gray lines represent phenomenological fits.}
\end{figure}

\subsection{\label{sec:level2}Heat Capacity}

Typical temperature scans of the specific heat at consecutive magnetic fields are shown in Fig.~\ref{fig:HC} for fields applied along the (a) \textbf{a} and (c) \textbf{b} directions. Sharp $\lambda$-type anomalies mark the antiferromagnetic long-range order (LRO) transition, where $T_N = 8$~K in zero field, consistent with susceptibility measurements. With increasing field, $T_N$ decreases, forming an LRO dome whose evolution agrees with the magnetization data in Fig.~\ref{fig:VSM}.

The specific heat probe allows for measurements below 2~K. Here, additional low-energy features appear in field scans as plotted in Figs.~\ref{fig:HC}(b, d). Along \textbf{b}, a pronounced hump around 8~T at 200~mK, which splits at higher temperatures, signals a field-induced gap closing and re-opening. A significantly weaker, but similar feature occurs along \textbf{a} at 200~mK in higher fields of 10~T. Its amplitude increases significantly upon warming, and it also splits into two components. Here, it seems the gap softens significantly but does not fully close due to the large suppression of heat capacity from 400 to 200~mK.

Additionally, in temperature scans at 8 and 14~T, along \textbf{b} [Fig.~\ref{fig:HC}(c)], a pronounced low-temperature upturn is observed. The enhancement is attributed to nuclear specific heat. This might contradict the fact that it only appears in regions where magnetic heat capacity is significant; however, it can be explained by nuclear degrees of freedom only being excited when hyperfine coupling of low-energy/softening magnons or phonons becomes relevant, as also observed in Nd$_3$BWO$_9$ \cite{NBWO}. This causes the upturn to be invisible in regions without low energy electronic/lattice degrees of freedom \cite{nuc1, nuc2}.

The specific heat measurements are plotted as false colorplot phase diagrams as a function of applied field and temperature in Figs.~\ref{fig:Phase diagram}(a, b) for an applied field along \textbf{a} and \textbf{b}, respectively.

\subsection{\label{sec:level2}Magnetostriction}

\begin{figure}[t]
\includegraphics[clip, trim=0.0cm 0.0cm 0.0cm 0.0cm ,width=\linewidth]{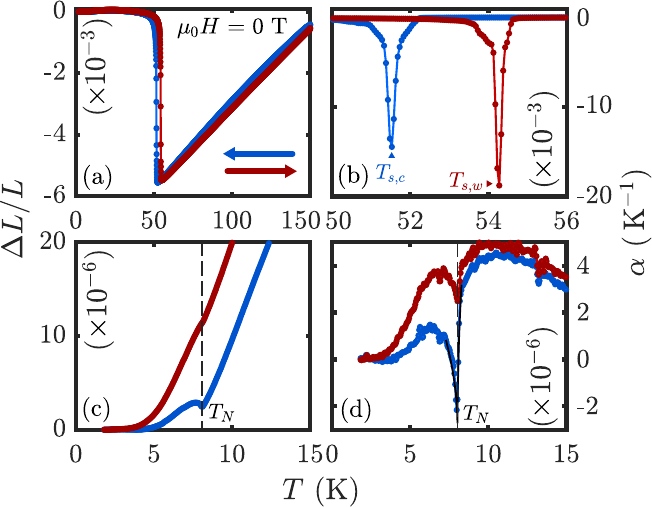}
\caption{\label{fig:dila} Sample dilation $\Delta L$ $||$ \textbf{c} in (a) and its (b) derivative (thermal expansion $\alpha$) against temperature measured while cooling (blue) and warming (red). (c) and (d) show zoomed-in versions below 15~K of the same quantities with adjusted scale. The black line on the cooling curve in (d) is a phenomenological fit to extract the peak position. Dashed lines indicate $T_N$.}
\end{figure}

\begin{figure}[t]
\includegraphics[clip, trim=0.0cm 0.0cm 0.0cm 0.0cm ,width=\linewidth]{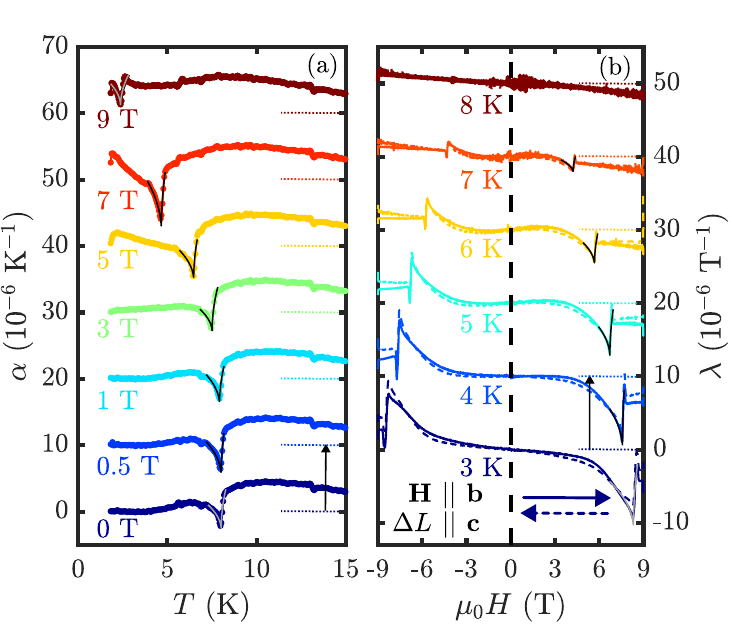}
\caption{\label{fig:Strictionc} (a) Thermal expansion from sample dilation measurements with $\Delta L$ $||$ \textbf{c} at several applied magnetic field along \textbf{b}. Consecutive curves are offset by 10$^{-5}$ K$^{-1}$. (b) magnetostriction coefficient at several temperatures. Up curves are solid lines, down sweeps dashed lines. Consecutive curves are offset by 10$^{-5}$ T$^{-1}$. Horizontal dotted lines indicate stacking baselines. Black and gray lines are phenomenological fits.}
\end{figure}

To probe any potential magneto-elastic coupling effects, we employ dilatometry measurements. In Fig.~\ref{fig:dila}(a), the relative length change $\Delta L/L$ along \textbf{c} is shown as a function of temperature upon cooling and warming in zero-field (ZF). Around $T$ $\sim$ 50 K, a sharp anomaly indicates the first-order structural phase transition. Its first-order nature is supported by the clear temperature hysteresis that is observed in the thermal expansion ($\alpha(T)= \frac{1}{L}\frac{\partial \Delta L}{\partial T}$) in Fig.~\ref{fig:dila}(b). A $\sim$3~K difference between $T_s$ upon cooling and warming is observed, where the warming transition point is consistent with susceptibility measurements in Fig.~\ref{fig:susceptibility}.

\begin{figure}[t]
\includegraphics[clip, trim=0.0cm 0.0cm 0.0cm 0.0cm ,width=\linewidth]{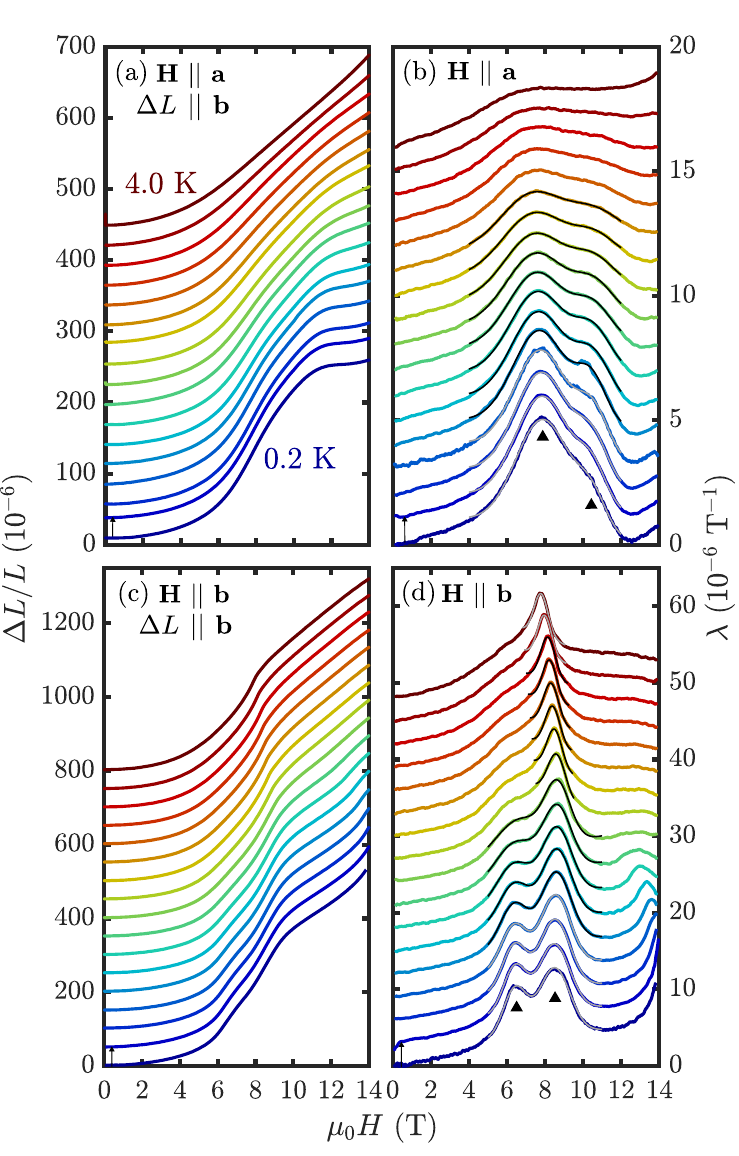}
\caption{\label{fig:Striction} Sample dilation $\Delta L$ $||$ \textbf{b} as a function of applied magnetic field along the (a) \textbf{a} direction and (c) \textbf{b} direction at several temperatures from 0.2 to 3~K in steps of 0.2~K and from 3 to 4~K in steps of 0.5~K. Consecutive curves are offset by (a) 30$\times$10$^{-6}$ and (c) 50$\times$10$^{-6}$. Plots (b) and (d) correspond to the magnetostriction coefficient along \textbf{a} and \textbf{b}, respectively. Consecutive curves are offset by (b) 10$^{-6}$~T$^{-1}$ and (d) 3$\times$10$^{-6}$~T$^{-1}$. Black arrows indicate anomalies at the lowest temperature, and black and gray lines are phenomenological fits.}
\end{figure}

In Figs.~\ref{fig:dila}(c) and (d), dilation and thermal expansion data are shown on a magnified scale below 15~K, in order to focus on the effect of the magnetic LRO transition on the length change around $T_N$. In the dilation, especially in the cooling curve, a clear anomaly is observed at $T_N$. In the thermal expansion this manifests in a strong negative peak, which is rather asymmetric. It increases sharply right above $T_N$ and gradually changes below. There is large hysteresis between the warming and cooling runs, where the magnitude of the peak also changes. Note that the thermal expansion here is roughly 5000 times smaller than at the structural transition. This is expected as the structural transition has a direct impact on the lattice, and is thus directly probed in thermal expansion measurements. On the other hand, the magnetic ordering transition is indirectly probed. It can be understood as an exchange-striction mechanism \cite{magnetostrict}. Exchange interactions depend on inter-atomic distances, so each bond acquires an extra stiffness contribution $\propto \langle\langle S_i \cdot S_j \rangle\rangle$ in the ordered state. Normally, the peak in thermal expansion at the transition temperature is positive (for example, in Ref. \cite{Nagl}) and the crystal shrinks to minimize the free energy. However, rather unusually, the thermal expansion has a negative peak here and thus expands.

Ehrenfest relations allow for an estimate of the initial pressure dependence of the transition temperature $\frac{\partial T_N}{\partial p} = V_mT_N\frac{\Delta \alpha}{\Delta C_p}$ \cite{Ehrenfest}. Where $p$ is the uni-axial pressure applied, $V_m$ is the molar volume, and $\Delta \alpha$ and $\Delta C_p$ are the heights of the anomaly in the specific heat and thermal expansion at $T_N$, respectively. Using the obtained data, we get $\frac{\partial T_N}{\partial p} \approx -0.1 $ K/GPa, meaning that $T_N$ is rather rigid against externally applied pressure.

\begin{figure}[t]
\includegraphics[clip, trim=0.0cm 0.0cm 0.0cm 0.0cm ,width=\linewidth]{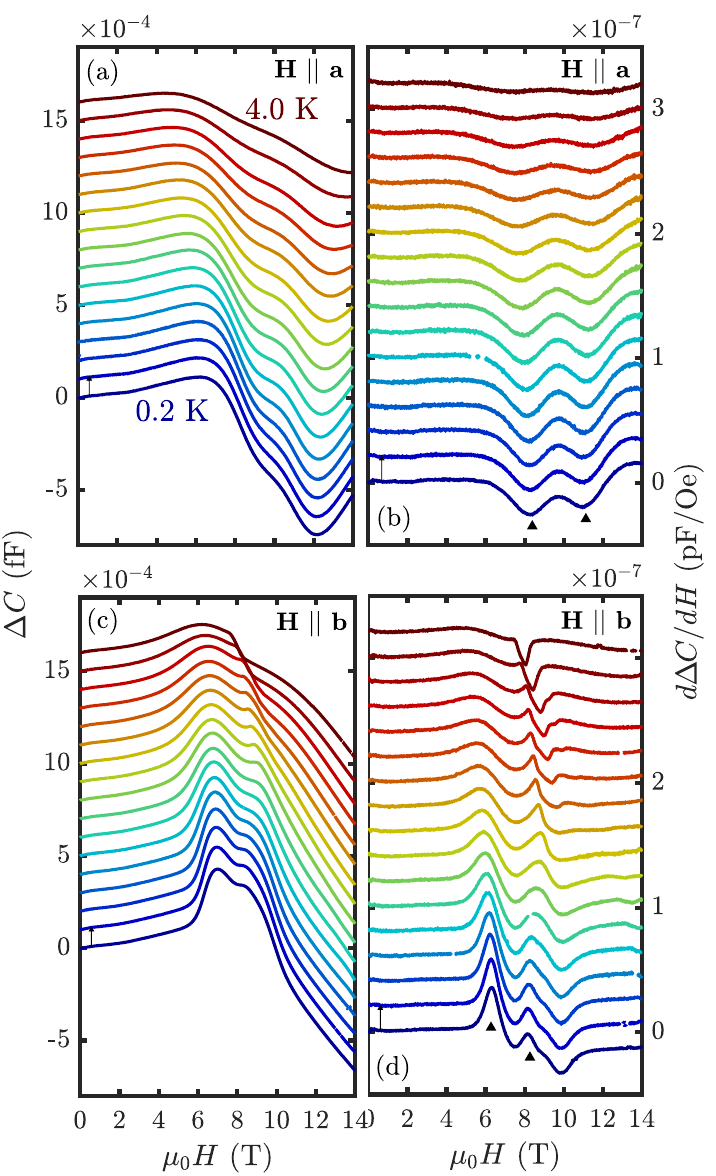}
\caption{\label{fig:Torque}Magnetic torque as a function of applied magnetic field along the (a) \textbf{a} direction and (c) \textbf{b} direction at several temperatures from 0.2 to 3~K in steps of 0.2~K and from 3 to 4~K in steps of 0.5~K. Consecutive curves are offset by 10$^{-4}$~fF. Plots (b) and (d) correspond to the derivative of the torque with respect to applied field along the \textbf{a} and \textbf{b} directions, respectively. Consecutive curves are offset by 0.2$\times$10$^{-4}$~pF/T.}
\end{figure}

\begin{figure*}[t]
\includegraphics[clip, trim=0.0cm 0.0cm 0.0cm 0.0cm ,width=\linewidth]{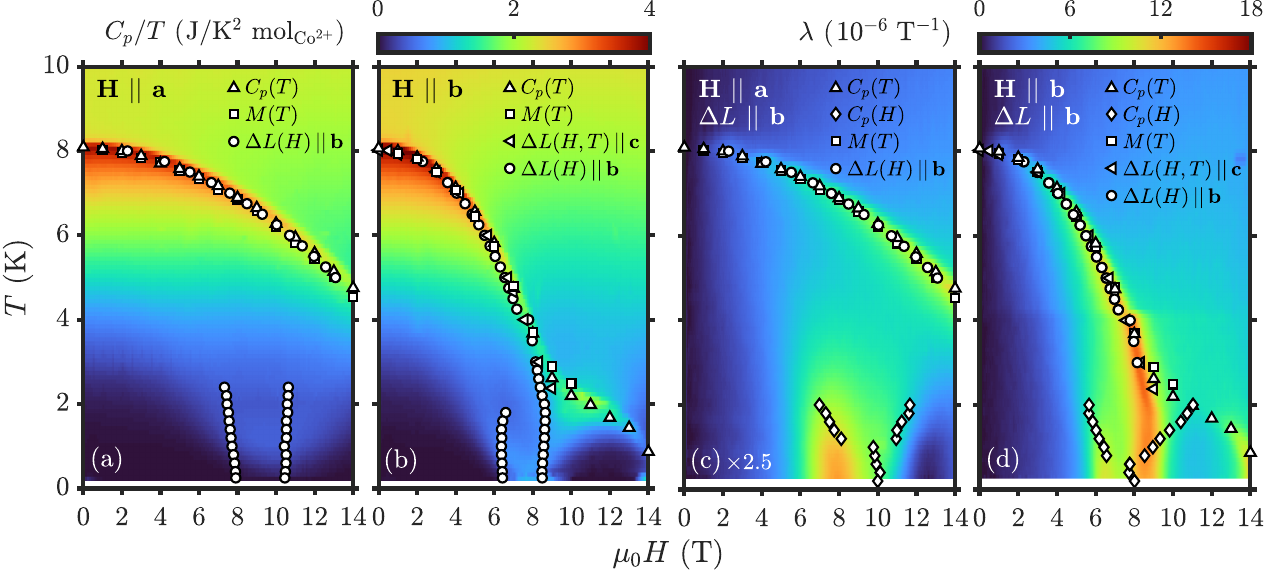}
\caption{\label{fig:Phase diagram}False-color plot of the specific heat measured in Cs$_2$CoI$_4$ in a magnetic field applied along the (a) \textbf{a} and (b) \textbf{b} directions. False-color plot of the magnetostriction coefficient $\lambda$ with applied magnetic field along the (c) \textbf{a} and (d) \textbf{b} directions. Symbols indicate the positions of anomalies as empirically fitted from the raw curves of the measurement probes discussed above.}
\end{figure*}

Cooling curves of the thermal expansion at several applied magnetic fields for \textbf{H} $||$ \textbf{b} are shown in Fig~\ref{fig:Strictionc}(a). $T_N$ decreases with increasing field while retaining the asymmetry. At 7~T, the low temperature tail of the peak pushes $\alpha(T)$ to finite values as $T$ $\rightarrow$ 0, which indicates the onset of the low temperature features as observed in specific heat. In addition, up and down sweeping magnetostriction coefficient ($\lambda(H) = \frac{1}{\mu_0L}\frac{\partial \Delta L}{\partial H}$) curves at several temperatures are visible in Fig~\ref{fig:Strictionc}(b). Just like in thermal expansion, the anomalies display an asymmetry towards higher fields. These measurements also show very few linear and quadratic components, and are mostly dominated by cubic-like tails ($H^3$) of the peaks at the phase boundaries. The asymmetric peaks at the phase boundary show only minor hysteresis. To extract the transition temperatures and fields, the curves were fitted phenomenologically. 

At lower temperatures, additional field-dependent measurements with \textbf{H} $||$ \textbf{a} and \textbf{b} of sample dilation $\Delta L$ $||$ \textbf{b} [Figs.~\ref{fig:Striction}(a, c)] and the magnetostriction coefficient [Figs.~\ref{fig:Striction}(b, d)] offer a different perspective on the low-temperature features observed in specific heat. At lowest temperatures of 200~mK, the scans reveal two well separated anomalies at 6.5 and 8.5~T along \textbf{b}, and two broader overlapping features at 8 and 10.5~T along \textbf{a}. These are roughly consistent with low temperature magnetization slope changes [Fig.~\ref{fig:LT_mag}], indicating potential magnetic phase transitions. With increasing temperature, the features gradually broaden. The two anomalies are well separated down to the lowest temperatures, in contrast with the apparent merging of the specific heat anomalies in both orientations [Fig.~\ref{fig:HC}]. To determine the peak positions, the anomalies have been phenomenologically fitted. Along \textbf{b}, the LRO dome appears at the highest fields and lowest temperatures and merges with the rightmost feature for increasing temperatures. 

False colorplots of the magnetostriction coefficient as a function of applied field and temperature are plotted in Figs.~\ref{fig:Phase diagram}(c) and (d) for an applied field along the \textbf{a} and \textbf{b} directions, respectively.

\subsection{\label{sec:level2}Magnetic torque}

Magnetic torque measurements performed below 4~K as a function of applied magnetic field are shown in Fig. \ref{fig:Torque}. Panels (a) and (b) display the torque signal and its field derivative for $\mathbf{H}$ $\mathbf{||}$ $\mathbf{a}$, while panels (c) and (d) show the corresponding data for $\mathbf{H}$ $\mathbf{||}$ $\mathbf{b}$. The quantity plotted is the field-induced change in capacitance, defined as $\Delta C = C(H) - C(H=0~\mathrm{T})$.

For $\mathbf{H}$ $\mathbf{||}$ $\mathbf{a}$, the 200~mK data exhibits strong changes in slope, which gradually broaden to higher temperatures. In the derivative [Fig. \ref{fig:Torque}(b)], these manifest in two broad continuous anomalies at 8 and 11~T. For $\mathbf{H}$ $\mathbf{||}$ $\mathbf{b}$, the 200 mK $\Delta C$ data show two broad anomalies centered near 6.5 T and 8.5 T, which diminish in amplitude as the temperature increases. For both orientations, the anomaly positions are consistent with magnetostriction coefficient data [Fig.~\ref{fig:Striction}]. Above roughly 2 K, along \textbf{b}, the LRO-dome becomes visible and appears as a sharp, discontinuous anomaly that merges with the rightmost hump. 

\begin{figure*}[t]
\includegraphics[clip, trim=0.0cm 0.0cm 0.0cm 0.0cm ,width=\textwidth]{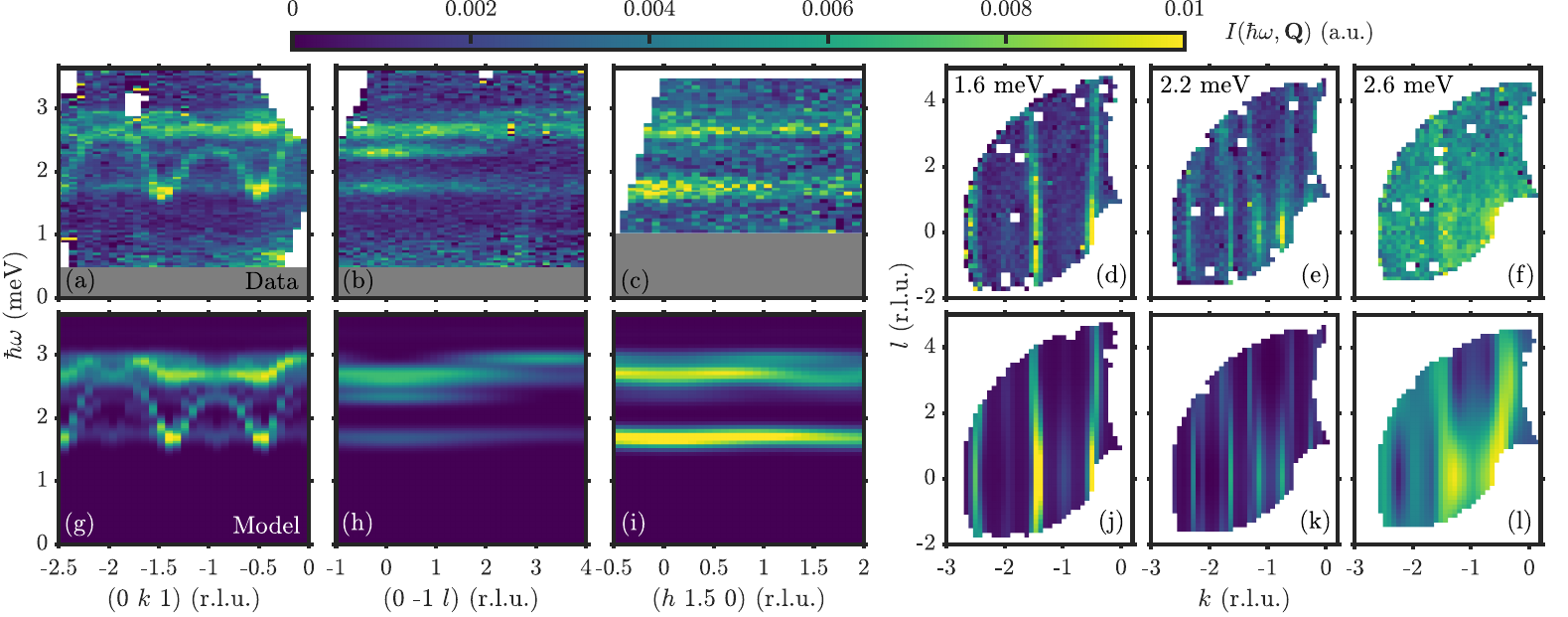}
\caption{\label{fig:INSZF} (a-c) show false color representations of energy momentum projections of measured inelastic neutron scattering intensity $I(\hbar\omega,\mathrm{\textbf{Q}})$ in Cs$_2$CoI$_4$ at $T$ = 100~mK and $\mu_0H$ = 0~T along three distinct reciprocal space directions. The data in (a) and (b) are integrated perpendicular to the scan direction in the $k$-$l$ plane in the range $\pm$0.2~r.l.u in $l$ and $\pm$0.1~r.l.u in $k$, respectively. Whereas the data in (c) is integrated perpendicular to the scan direction in the $h$-$k$ plane in the range $\pm$0.1~r.l.u in $k$. Since (c) is from another instrument, the data were normalized to match the intensity of (a) and (b). (d-f) are false color representations of measured constant energy slices, where the integration range in energy is $\pm$0.05~meV. The gray boxes in (a) and (b) mask out the elastic line and in (c) an area that was not measured. White patches are masked out spurions. The figures in the second row, (g-l), show SU(4) spin wave theory calculations based on the minimal Hamiltonian. These are directly associated with the measured data in the same column as the row above.}
\end{figure*}

\subsection{\label{sec:level2}Magnetic phase diagram}

Anomalies from magnetization, specific heat, thermal expansion, and magnetostriction coefficient were fitted phenomenologically and are superimposed as symbols in the $H-T$ magnetic phase diagrams of Fig.~\ref{fig:Phase diagram}. The phase diagrams consist of false color-plots of specific heat as shown in Figs.~\ref{fig:Phase diagram}(a) and (b) with applied field along the crystallographic \textbf{a} and \textbf{b} directions, respectively. In addition, false color-plots of magnetostriction coefficient data are shown in Figs.~\ref{fig:Phase diagram}(c) and (d) with applied field along the crystallographic \textbf{a} and \textbf{b} directions, respectively. 

The LRO dome is consistent across all probes, whereas the low-temperature features show probe-dependent behavior in both orientations. As discussed above, two anomalies persist to the lowest temperature in magnetostriction data, while specific heat anomalies appear to merge. To emphasize the different behavior of these probes at low temperatures, the anomalies from magnetostriction coefficient have been exclusively plotted on the specific heat phase diagram [Figs.~\ref{fig:Phase diagram}(a) and (b)], and the low temperature anomalies from specific heat against field have been only plotted on the magnetostriction phase diagrams [Figs.~\ref{fig:Phase diagram}(c) and (d)]. It is evident that the field position of the merged anomalies in specific heat seems to be consistent with one of the two anomalies observed in striction. This indicates that most likely not all field induced phase transitions are probed with specific heat. Additional phase transitions appearing in striction measurements compared to heat capacity are not unique to this material, as it was also observed in (CD$_3$ND$_3$)$_2$NaRuCl$_6$ \cite{Nagl}.

\subsection{\label{sec:level2}Neutron Spectroscopy}

\begin{table}[b]
\caption{\label{tab:interactions}
Parameters of a minimal Heisenberg Hamiltonian for Cs$_2$CoI$_4$.}
\begin{ruledtabular}
\setlength{\tabcolsep}{6pt} 
\begin{tabular}{lc}
Parameter & Value (meV) \\
\hline
$J_A$   &  0.37 \\
$J'_A$  &  0.15 \\
$D_A$   &  2.00 \\
$E_A$   &  0.62 \\
\hline
$J_B$   &  0.20 \\
$J'_B$  &  0.08 \\
$D_B$   &  0.30 \\
$E_B$   &   0.90 \\
\end{tabular}
\end{ruledtabular}
\end{table}

Inelastic neutron scattering measurements were performed on Cs$_2$CoI$_4$ at CAMEA (PSI) and HODACA (JRR-3). At CAMEA, the sample was aligned in the $(0,k,l)$ horizontal scattering plane and measured at $T$ = 100~mK in magnetic fields $\mu_0H = 0$, 3, and 6~T applied along \textbf{a}. Complementary ZF data were collected at HODACA at $T$ = 700~mK with the sample aligned in the $(h,k,0)$ plane. No background was subtracted.

Representative energy–momentum projections and constant energy cuts of the ZF spectrum are shown in \ref{fig:INSZF}(a-f). The excitation spectrum is gapped and considerably 1D, dispersing predominantly along the $k$ direction, while remaining flat within experimental resolution along $h$ and $l$. The constant energy cut around the gap energy of 1.6 meV [Fig.~\ref{fig:INSZF}(d)] does show a weak modulation along $l$. The excitation spectrum can be roughly divided in three sectors. The dominant dispersive mode (between 1.6 and 2.5~meV) emerges from the commensurate wave vector $\mathbf{Q}=(0,1/2,0)$ and is symmetrically skewed towards the Brillouin zone center at (0, -1, 1). In addition, a flat excitation branch is observed at 1.75~meV. Furthermore, between 2.4 - 3~meV, another broader band of intensity emerges. It appears to contain a slight modulation/multiple modes which cannot be resolved within the instrumental resolution.

\begin{figure}[t]
\includegraphics[clip, trim=0.0cm 0.0cm 0.0cm 0.0cm ,width=\linewidth]{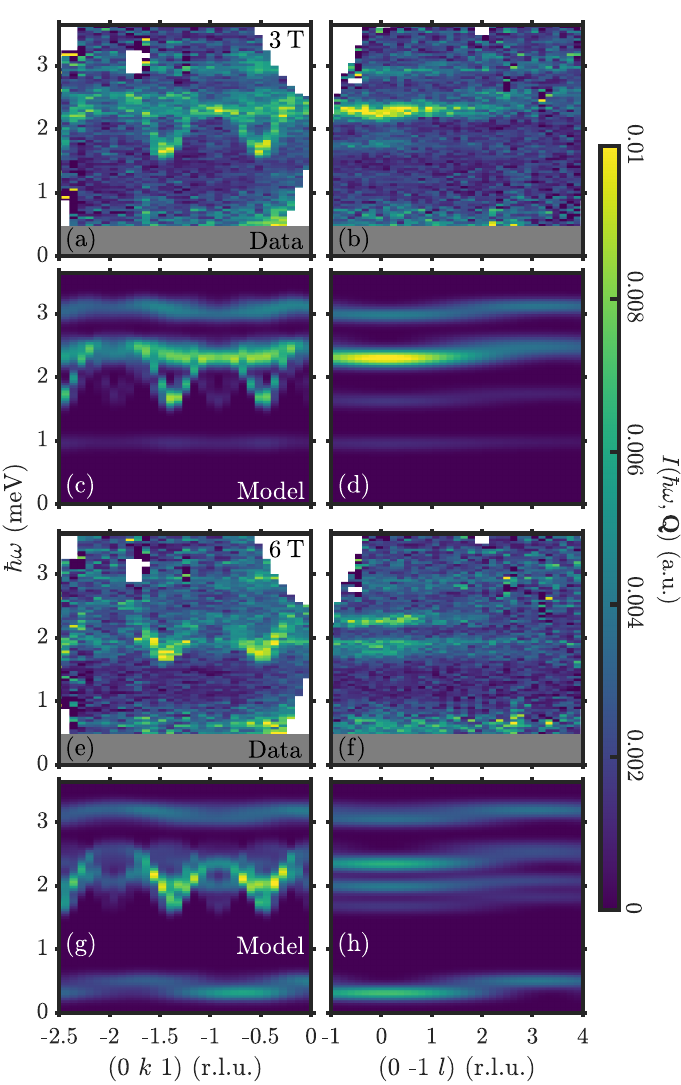}
\caption{\label{fig:INSField} False color representations of energy momentum projections of measured inelastic neutron scattering intensity $I(\hbar\omega,\mathrm{\textbf{Q}})$ in Cs$_2$CoI$_4$ at $T$ = 100~mK. The data in subplots (a) and (b) were measured at $\mu_0H$ = 3~T and  data in (e) and (f) at $\mu_0H$ = 6~T, both for applied field along \textbf{a}. The data are integrated perpendicular to the scan direction in the $k$-$l$ plane in the range $\pm$0.2~r.l.u in $l$ for $k$-scans and $\pm$0.1~r.l.u in $k$ for $l$-scans, respectively. Figures (c), (d), (g), and (h) show SU(4) spin wave theory calculations based on the minimal Hamiltonian. These are directly associated with the measured data in the same column as the row above.}
\end{figure}

Exemplary energy-momentum projections along $k$ and $l$ in applied magnetic fields of $\mu_0H$ = 3 and 6~T along the crystallographic \textbf{a} direction are illustrated in Figs.~\ref{fig:INSField}(a,b) and (e,f), respectively. The main dispersive branch remains largely unaffected up to 6~T. In contrast, the flatter excitation branches exhibit pronounced field dependence. At 3~T, a clear Zeeman splitting of the upper flat mode is observed, while at 6~T the split modes become weak and difficult to distinguish from background. Additional low-energy intensity appears below 1 meV at 6~T.

\subsection{\label{sec:ModelHamiltonian}Model Hamiltonian and INS analysis}

\begin{figure}[t]
\includegraphics[clip, trim=0.0cm 0.0cm 0.0cm 0.0cm ,width= 8cm]{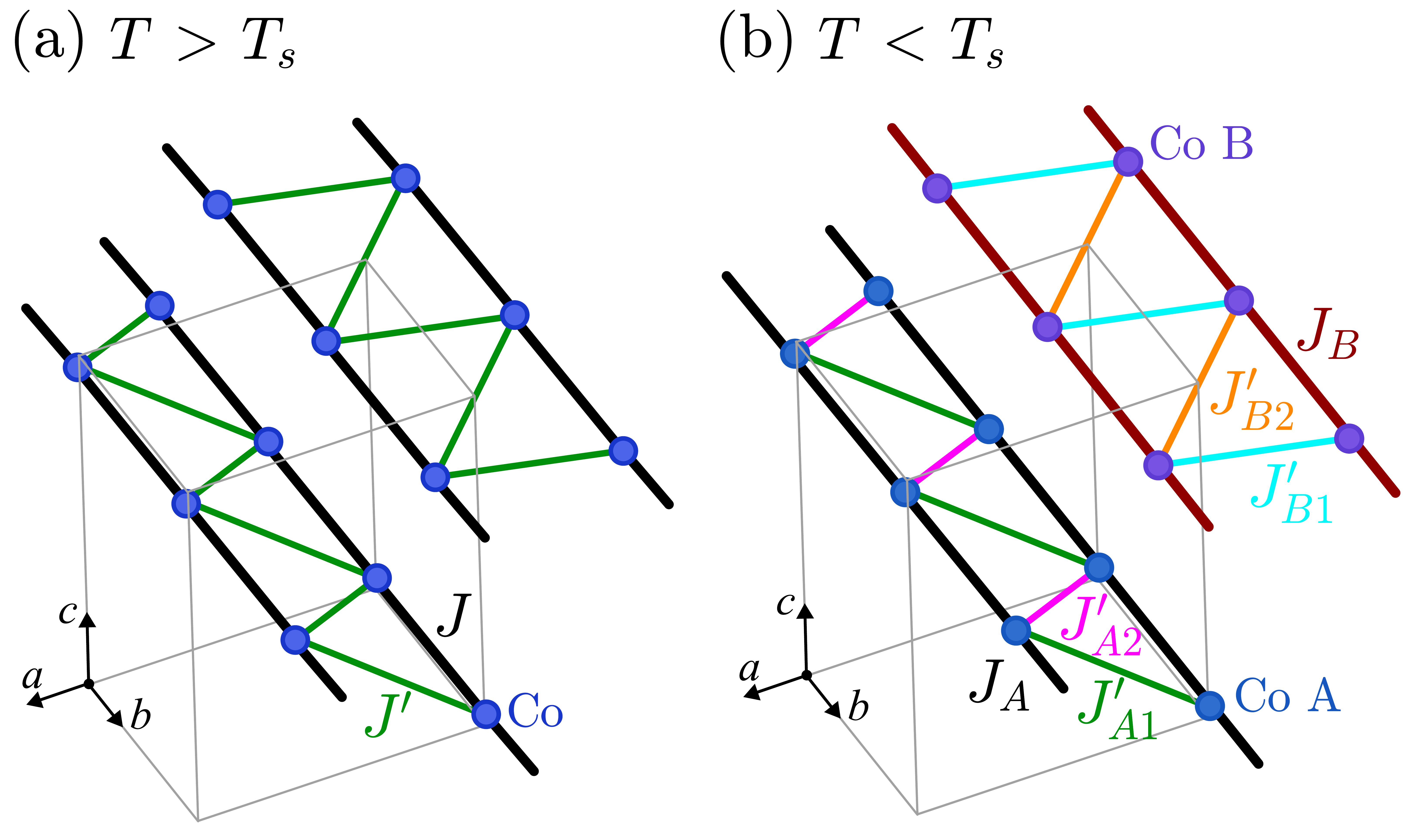}
\caption{\label{fig:exchangepaths} Schematic overview of the allowed exchange interactions in the zig-zag ladders at (a) room temperature and (b) low temperature (note that the orthorhombic convention is used here to label the axes).}
\end{figure}

\begin{figure}[b]
\includegraphics[clip, trim=0.0cm 0.0cm 0.0cm 0.0cm ,width= 8cm]{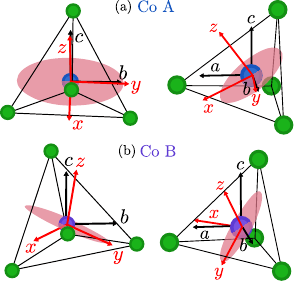}
\caption{\label{fig:pointcharge} Schematic overview of the local axes diagonalizing the single-ion anisotropy tensor as obtained from point-charge calculations for (a) Co A and (b) Co B. Left is the view from the $bc$ plane and right an almost orthogonal orientation. The local axes are indicated in red by $x$, $y$ and $z$. Transparent red planes indicate the single-ion anisotropy ellipses based on values in Table~\ref{tab:interactions}.}
\end{figure}

With the goal to determine a minimal model spin Hamiltonian for Cs$_2$CoI$_4$, the inelastic neutron scattering data were analyzed using SU(4) spin wave theory (SWT) computations performed using the \texttt{SUNNY.JL} software package \cite{sunny}. We followed an approach similar to that previously used for Cs$_2$CoBr$_4$ and Cs$_2$RuO$_4$ \cite{Facheris2024, CRO}.

Assuming Heisenberg interactions with additional single-ion anisotropy, the reduced symmetry of the low-temperature crystal structure allows for sixteen nearest neighbor exchange interactions and twelve single-ion anisotropy parameters. This renders a comprehensive quantitative fit of the excitation spectrum impractical. Instead, we adopt a physically motivated minimal Hamiltonian that captures the essential features of the observed spectrum.

Motivated by the pronounced 1D character of the excitations and the similarity of the ZF spectrum (excluding the Zeeman ladder), the model is restricted to zigzag ladder geometries analogous to those established in Cs$_2$CoBr$_4$ \cite{Facheris2024}, as shown in Fig.~\ref{fig:exchangepaths}(a). Below the structural transition, the two equivalent ladders per unit cell become inequivalent and are denoted ladder A and ladder B, constructed by the inequivalent Co~A and Co~B, respectively. These ladders and their allowed interactions are illustrated in Fig.~\ref{fig:exchangepaths}(b). Each ladder is characterized by an intrachain exchange $J_{A,B}$ and two zigzag couplings $J'_{A1,B1}$ and $J'_{A2,B2}$. The symmetry-inequivalent zigzag paths within a ladder are assumed to be equal to limit the parameter space and are indicated by $J'_{A,B}$. 

Single-ion anisotropy is included in its most general diagonal form in a local coordinate frame for each ion:

\begin{equation}
{\cal H}_{SI} = -E_{A,B}\hat{S}_y^2 + D_{A,B}\hat{S}_z^2.
\label{eq:SI}
\end{equation}

To reduce the number of parameters, the local frames that diagonalize the single-ion anisotropy tensors are required. Here, these are estimated from point-charge crystal-field calculations, which provide physically reasonable orientations of the anisotropy axes for the two inequivalent Co$^{2+}$ sites. These were performed using PyCrystalField \cite{PCF}. We confirmed that this approach reproduces the known anisotropy axes in related compound Cs$_2$CoBr$_4$ \citep{Povarov2020} and is therefore used here to constrain the anisotropy sector of Cs$_2$CoI$_4$.

The calculated local coordinate systems of the two Co$^{2+}$ ions in Cs$_2$CoI$_4$ are schematically illustrated in Fig.~\ref{fig:pointcharge} with red $xyz$ axes. The local frame on Co~A [Fig.~\ref{fig:pointcharge}(a)] stays similar to Cs$_2$CoBr$_4$ as the local environment is not distorted drastically. The distortion is quite severe for Co~B [Fig.~\ref{fig:pointcharge}(b)], and thus also projected in the change in the local axes.

In equation \ref{eq:SI}, the $z$ axis defines the normal to an easy plane with strength $D_{A,B}$. One perpendicular axis $y$ defines an in-plane easy axis component $E_{A,B}$. The $y$ and $z$ axes and single-ion anisotropy parameters will be chosen per inequivalent ion to obtain the best match of the resulting simulation with the measured neutron spectrum. 

The final minimal model is based on the assumptions mentioned above. Nonetheless, there are still ten parameters to be determined (including $g$-factors along \textbf{a} for both ions) and four local axes ($y$ and $z$ on each ion) to be chosen. Under these conditions, it is difficult to introduce a quantitative measure of "goodness of fit". For these reasons, the analysis is not a fit. Instead, a trial and error approach was utilized to select an ad hoc set of parameters in the proposed minimal model to reproduce the main features of the data. These include periodicities, bandwidths, gaps, and qualitative obvious intensity modulations. To enable direct comparison to the experiment, the computation assumed a Gaussian energy resolution of $\sigma$ = 0.19~meV standard deviation to roughly match those of the instruments. The corresponding simulated spectra are shown alongside the data in Figs.~\ref{fig:INSZF} and \ref{fig:INSField}. The resulting parameter set is summarized in Table~\ref{tab:interactions}.

In ZF, the model reproduces the experimental spectrum remarkably well. The dominant dispersive mode originates from ladder A, with its bandwidth set primarily by $J_A$ and its skewness and weak $l$ modulation controlled by $J'_A$. The excitation gap is determined by a combination of $J_A$, $D_A$, and $E_A$. A higher-energy crystal-field excitation (CEF) with moderate intensity is predicted by the model but lies outside the experimentally accessible energy window. Since the distance to this CEF mode was not measured, and there is a degree of co-dependence on the parameters, a set of $J_A$, $E_A$, and $D_A$ will give similar results. Here, we propose one of these solutions, where the CEF sector is high enough in energy to be outside of the measured window in zero and applied field. 

The flatter excitation branches originate from ladder B. Their reduced bandwidth directly reflects the smaller exchange scale $J_B$, while the skewness of the highest mode is set by $J'_B$, which is difficult to resolve from measurements. The relative positions of the two flat modes require a dominant easy-axis anisotropy component, $E_B > D_B$.

In ZF, only the magnitude of the single-ion anisotropy parameters are of relevance. The orientation/selection of the easy and hard axis from the local $xyz$ axes play only a minor role, besides small polarization factor effects. The orientation of the local axes becomes essential to reproduce the field dependence of the spectrum. For ladder A, the resulting single ion anisotropy ellipsoid is illustrated in Fig.~\ref{fig:pointcharge}(a). Similar to Cs$_2$CoBr$_4$, the easy axis $y$ lies predominantly along \textbf{b} and hard axis $z$ is rotated approximately 45$^{\circ}$ away from the \textbf{a} axis in the $ac$ plane. The easy \textbf{b} axis renders ladder A's excitations largely insensitive to a field applied along \textbf{a}, as observed in the data. In contrast, the flat branches from ladder B display clear Zeeman splitting. Thus, the easy-axis $y$ is chosen along a local axis with a sizable projection along \textbf{a}. The axis with the smallest \textbf{a} component has been chosen as the axis normal to the easy plane $z$. The ellipsoid for ladder B is shown in Fig.~\ref{fig:pointcharge}(b).

The $g$-tensors for both ladders are only diagonal in the correct local axis frame. Since the field is solely applied along the \textbf{a} direction, an effective $g$-factor along that direction for both ladders was used. They were selected to optimally reproduce the splitting observed in applied field and their average value to match $g_a$ from the high field magnetization data. This results in $g_{\mathrm{eff, A}}$ = 2.2 and $g_{\mathrm{eff, B}}$ = 2.8. 

\section{\label{sec:level1}Discussion}
\begin{figure}[b]
\includegraphics[clip, trim=0.0cm 0.0cm 0.0cm 0.0cm ,width= 8cm]{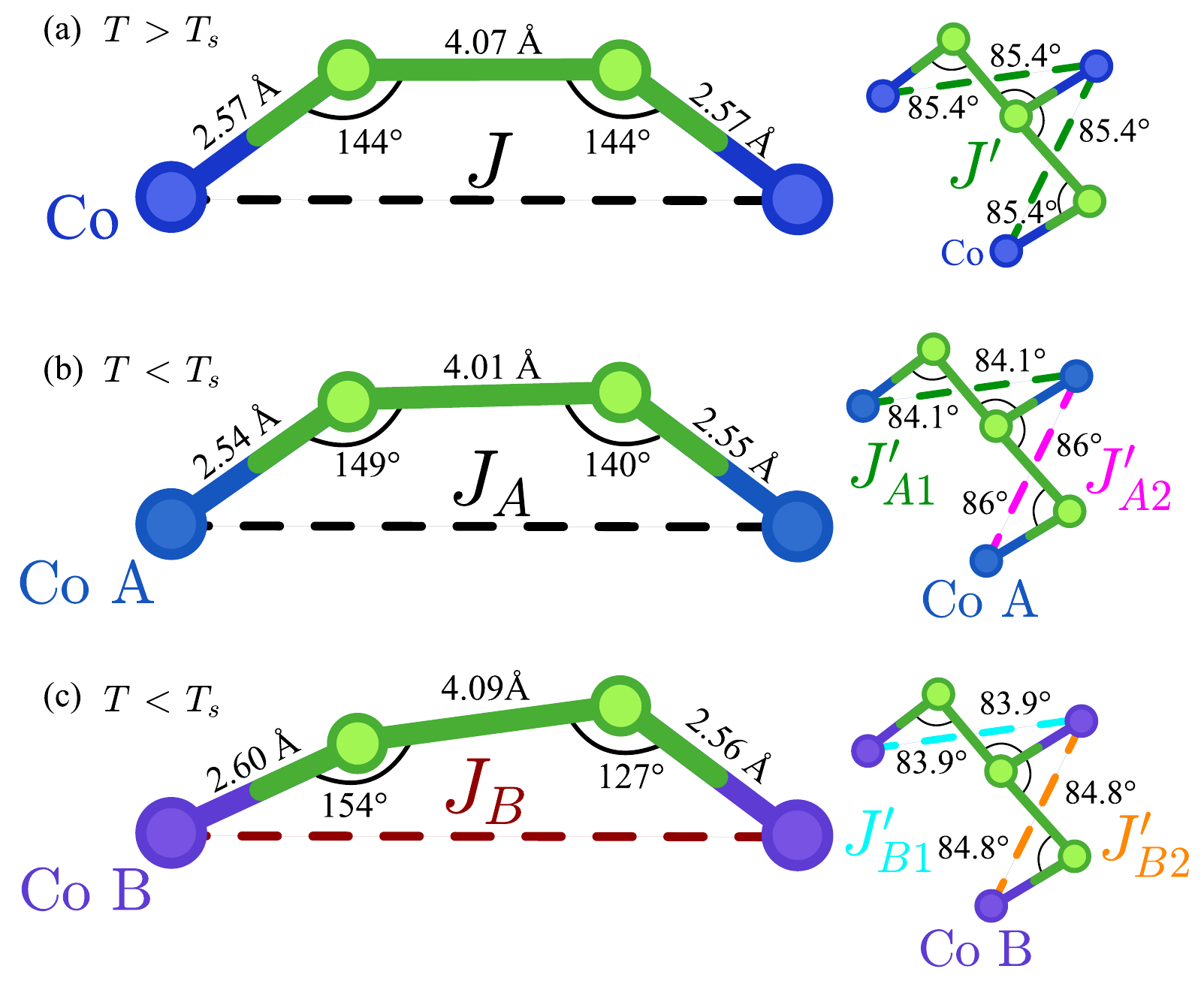}
\caption{\label{fig:exchange_paths_super} Schematic overview of the chain (left column) and zig-zag (right column) Co-I-I-Co super-super exchange paths with associated distances and angles at (a) room temperature structure, (b) low temperature structure in ladder A, and (c) low temperature structure in ladder B.}
\end{figure}

\subsection{Model Hamiltonian}

In this work, we propose a minimal magnetic Hamiltonian for Cs$_2$CoI$_4$ consisting of two zigzag ladders, Ladder A and Ladder B. Ladder A is formed by Co A sites and Ladder B by Co B sites, where Co A and Co B are symmetrically inequivalent Co$^{2+}$ ions. Owing to the reduced symmetry of the low-temperature crystal structure, the number of symmetry-allowed interactions is large, precluding a unique determination of all model parameters. The Hamiltonian presented here should therefore be regarded as a physically motivated effective description rather than a definitive microscopic model.

The exchange interactions and single-ion anisotropies differ substantially between the two ladders. In particular, the intrachain exchange along the \textbf{b} direction in ladder A, $J_A$, is approximately twice as large as the corresponding interaction $J_B$ in ladder B. Both interactions arise from Co–I–I–Co super–super-exchange pathways. While the associated bond angles remain largely unchanged in ladder A across the structural transition, ladder B undergoes a pronounced distortion, as illustrated in Fig.~\ref{fig:exchange_paths_super}(a-c, left). This is expected to strongly modify orbital hybridization and reduce the effective exchange strength.

For the zigzag interactions, we assume that the two symmetry-allowed exchange paths \textit{within} each ladder can be represented by a single effective coupling. This approximation is justified by the comparatively small structural distortions affecting these paths and the similar bond angles observed for ladders A and B, as illustrated in Fig.~\ref{fig:exchange_paths_super}(a-c, right). Consequently, the effective zigzag exchanges $J'_A$ and $J'_B$ are found to be closer in magnitude than the chain-like interactions.

The minimal Hamiltonian provides a satisfactory description of the ZF excitation spectrum, but discrepancies emerge at higher magnetic fields. While the spectrum of ladder A remains in reasonable agreement up to 6~T, and the Zeeman splitting of the flatter modes associated with ladder B is captured at 3~T, the agreement clearly deteriorates at 6~T. For example, the flat modes around 0.8~meV and 2.9~meV in the data are slightly offset compared to the simulation. This behavior reflects the obvious limitations of the minimal model. In particular, the local axes entering the single-ion Hamiltonian are derived from a point-charge calculation and represent an approximate description. Small deviations in the orientation of these axes, as well as uncertainties in the anisotropy tensor and $g$-tensor components, can significantly affect the splitting of branches in the high-field excitation spectrum. Moreover, these parameters are not independent, implying that multiple parameter sets may yield comparable agreement with experiment. A follow-up inelastic neutron scattering experiment probing higher energy transfers could further constrain the coupled parameters by accessing the CEF level predicted by the full SU(4) model, but beyond the current energy range.

Finally, the model treats the two ladders as decoupled, rendering the system effectively 1D. As a result, interaction-induced renormalizations are expected at ZF analogous to those reported in related quasi-1D systems such as Cs$_2$CoBr$_4$ \cite{Facheris2024}. In addition, the disordered ground state predicted by the 1D model contrasts with the experimentally observed long-range order below $T_N = 8$~K. This ordering must originate from residual inter-ladder couplings, which are neglected here but may be significant, given the relatively high N\'eel temperature, and play a role in the field evolution of the spectrum.

\subsection{Magnetic phase diagram}

For a classical antiferromagnetic zigzag ladder, the ground state is an incommensurate spiral, approaching a collinear state in the limit $4J \gg J'$ \cite{zigzag}. In Cs$_2$CoI$_4$, no evidence for a spiral phase is observed. Instead, the phase diagram exhibits a gapped state in ZF, and the excitation spectrum emerges from commensurate wave vectors \textbf{Q} = (0, 1/2, 0), consistent with a collinear magnetic structure. The spiral instability is suppressed by the relatively strong easy-axis anisotropies, and the most likely ground state is an antiferromagnetic Néel phase with spins aligned along the local easy axes of each ladder, in agreement with the ground state from SWT calculations.

A striking feature of the magnetic phase diagram is the pronounced softening of the excitation gap for magnetic fields applied along the \textbf{a} and \textbf{b} directions. For fields along \textbf{b}, the gap appears to close at a single transition at low temperatures, separating two gapped phases within the ordered regime. In contrast, for fields along \textbf{a}, the gap softens but seems to remain finite. The minimal model allows these trends to be qualitatively understood from the anisotropy structure of the two ladders. Ladder A has a dominant easy-axis component along \textbf{b}, such that a field along this direction efficiently drives a Zeeman-induced gap closure. Ladder B, in contrast, has an easy axis with a substantial but non-collinear component along \textbf{a}, resulting in a gap softening without complete closure.

The gap-softening feature bears a superficial resemblance to the spin-flop–like transition and associated quantum critical point observed in Cs$_2$RuO$_4$ \cite{CRO}. However, the situation in Cs$_2$CoI$_4$ is fundamentally different. Within each ladder, the anisotropy axes are uniform by symmetry, precluding a continuous spin-flop transition, and thermodynamic measurements indicate multiple anomalies for both field directions, in contrast to the singular anomalies observed in Cs$_2$RuO$_4$.

Magnetostriction and torque measurements reveal at least two distinct anomalies persisting to low temperatures for fields applied along either crystallographic direction, whereas specific heat measurements suggest a single transition at the lowest temperatures and a splitting only at elevated temperatures. This discrepancy may indicate that the specific heat associated with the gap softening dominates over signatures of potentially split phase transitions at low temperature. Another explanation would be that specific heat is largely insensitive to one of the two field-induced phase transitions.

For an isolated zigzag ladder with strong easy-plane and weaker easy-axis anisotropy (such as ladder A), multiple field-induced transitions are expected for fields applied along the easy axis, including magnetization plateaus and spin-flop phases, as observed in Cs$_2$CoBr$_4$ \cite{Facheris2024}. In Cs$_2$CoI$_4$, such behavior is not observed and may be masked by the simultaneous response of both ladders and potential residual inter-ladder couplings, which are neglected in the minimal model but may significantly affect the phase evolution. 

Although the resulting phase diagram is highly intriguing, a complete understanding of the field-induced phases in Cs$_2$CoI$_4$ remains beyond the scope of the present minimal model. Single crystal neutron diffraction measurements would be crucial to directly determine the magnetic structures; however, the possible solutions are barely constrained by symmetry. Additionally, it requires careful treatment of multiple crystallographic domains arising from the structural transition, or the use of an in-field slow-cooling protocol as in the Appendix, compatible with neutron diffraction.

\section{\label{sec:level1}Conclusion}
The $S$ = 3/2 antiferromagnet Cs$_2$CoI$_4$ goes through a structural phase transition to a crystal structure with low symmetry. This causes the magnetic Hamiltonian to be heavily over-parametrized. We propose a minimal model constructed of two inequivalent zigzag ladders that reasonably reproduces the inelastic neutron spectra.  

\section*{\label{sec:ack}Acknowledgements}
The work at ETH Z\"urich is supported by a MINT grant of the Swiss National Science Foundation. This work is partly based on experiments performed at the Swiss spallation neutron source SINQ, Paul Scherrer Institute, Villigen, Switzerland, where beam time was allocated at CAMEA (ID: 20222515) and HRPT (ID: 20220314). We also acknowledge the support of the HLD at HZDR, member of the European Magnetic Field Laboratory (EMFL), and the W\"{u}rzburg-Dresden Cluster of Excellence $ctd.qmat$ -- Complexity, Topology and Dynamics in Quantum Matter (EXC 2147, project No. 390858490). We acknowledge Diamond Light Source for time on I19 EH1 under proposal CY37825.

\section*{\label{sec:appA}Appendix: High field slow cooling procedure}

\begin{figure}[b]
\includegraphics[clip, trim=0.0cm 0.0cm 0.0cm 0.0cm ,width=\linewidth]{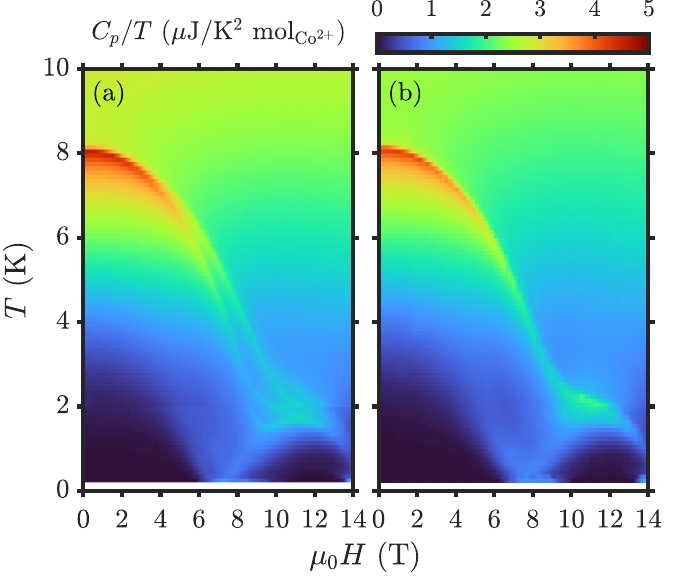}
\caption{\label{fig:twin}False-color plot of the specific heat measured in Cs$_2$CoI$_4$ in a magnetic field applied along the \textbf{b} direction measured while (a) fast cooling (10~K/s) over $T_s$ in zero-field and (b) slow cooling (0.1~K/s) over $T_s$ in an applied field of 14~T.}
\end{figure}

When measuring the thermodynamic properties of Cs$_2$CoI$_4$, multiple features were observed at moderate to high magnetic fields, as illustrated for the specific heat in Fig.~\ref{fig:twin}(a). The LRO phase boundary splits above $\sim$5~T and further branches above $\sim$8~T, indicative of multi-domain effects. These originate from the structural phase transition, which splits the crystal into multiple domains. The number of symmetry-allowed domains is given by $n = |G|/|F|$, where $|G|$ and $|F|$ are the number of symmetry operations in the high and low temperature symmetry group, respectively \cite{Domain}, yielding four possible domains for the present transition.

Due to the change in lattice angles between the orthorhombic and triclinic structures, the domains are slightly misaligned with respect to the applied magnetic field. Upon zero-field cooling, all domains are populated, resulting in multiple apparent phase transitions.

Multi-domain effects can be largely suppressed by slowly cooling (0.1~K/min) through the structural transition in a magnetic field of 14~T. Although the reason is not fully understood, the likely explanation is that the field selects a preferred domain and yields as clean a phase diagram as possible, as shown in Fig.~\ref{fig:twin}(b). This domain selection is most likely driven by magnetoelastic coupling. This protocol was applied to all thermodynamic measurements, except pulsed-field magnetization, where it was technically unfeasible but still allowed reliable estimates of the saturation field and magnetization.

\bibliography{bibliography.bib}

\end{document}